\documentclass[conference]{IEEEtran}
\IEEEoverridecommandlockouts

\usepackage{cite}
\usepackage{amsmath,amssymb,amsfonts,amsthm}
\usepackage{algorithm}
\usepackage{algorithmic}
\usepackage{graphicx}
\usepackage{textcomp}
\usepackage{xcolor}
\usepackage{hyperref}
\usepackage{booktabs}
\usepackage{balance}

\hypersetup{
    colorlinks=true,
    linkcolor=blue,
    filecolor=magenta,      
    urlcolor=cyan,
    citecolor=blue,
}

\newtheorem{definition}{Definition}

\begin{document}

\title{IslandRun: Privacy-Aware Multi-Objective Orchestration for Distributed AI Inference}

\author{\IEEEauthorblockN{Bala Siva Sai Akhil Malepati\thanks{Portions of this work are subject to Indian Provisional Patent Application No. 202541108678, filed November 10, 2025.}}
\IEEEauthorblockA{Email: saiakhil2012@yahoo.com}
}

\maketitle

\begin{abstract}
Modern AI inference faces an irreducible tension: no single computational resource simultaneously maximizes performance, preserves privacy, minimizes cost, and maintains trust. Existing orchestration frameworks optimize single dimensions (Kubernetes prioritizes latency, federated learning preserves privacy, edge computing reduces network distance), creating solutions that struggle under real-world heterogeneity. We present \textbf{IslandRun}, a multi-objective orchestration system that treats computational resources as autonomous ``islands'' spanning personal devices, private edge servers, and public cloud. Our key insights: (1) \emph{request-level heterogeneity demands policy-constrained multi-objective optimization}, (2) \emph{data locality enables routing compute to data rather than data to compute}, and (3) \emph{typed placeholder sanitization preserves context semantics across trust boundaries}. IslandRun introduces agent-based routing, tiered island groups with differential trust, and reversible anonymization. This establishes a new paradigm for privacy-aware, decentralized inference orchestration across heterogeneous personal computing ecosystems.
\end{abstract}

\begin{IEEEkeywords}
edge computing, privacy-preserving inference, multi-objective optimization, LLM orchestration, trust-aware routing, distributed AI
\end{IEEEkeywords}

\section{Introduction}

\subsection{The Multi-Objective Inference Dilemma}

The commoditization of large language models (LLMs) and diffusion-based generative AI has exposed a fundamental tension in inference deployment: \emph{no single computational island satisfies all operational objectives simultaneously}. Cloud providers offer unlimited scale but violate data sovereignty constraints; local devices preserve privacy but exhaust resources under load; edge servers balance proximity and capacity but introduce operational complexity. This heterogeneity is not a temporary artifact of immature infrastructure. It is an intrinsic property of the privacy, performance, and cost tradeoff space~\cite{dwork2014algorithmic,kumar2010cloud}.

Consider four real-world scenarios illustrating this heterogeneity:

\textbf{Scenario 1: Personal Island Group}  
A user's devices (laptop, mobile phone, smart TV, car infotainment) form a \emph{trusted island group}. When working on the laptop, a code completion request uses the laptop's GPU. When driving, the same conversation continues on the car's system, but routes to the user's home laptop (if available) or mobile hotspot for privacy-sensitive queries, falling back to cloud only for non-sensitive tasks.

\textbf{Scenario 2: Dynamic Resource Sharing}  
Two friends hiking in a remote area need to process photos with AI enhancement. Friend A's phone has low battery but strong cellular signal; Friend B's phone has high battery but weak signal. IslandRun automatically detects this imbalance and routes Friend A's inference requests to Friend B's device via Bluetooth mesh, preserving both users' privacy while optimizing battery consumption.

\textbf{Scenario 3: Data Locality Routing}  
A legal firm has vectorized 10TB of case law on their private server. Instead of uploading documents to cloud LLMs, IslandRun routes queries \emph{to where the data already exists}, executing RAG inference on the private server where embeddings are stored. This inverts traditional ``data to compute'' into ``compute to data.''

\textbf{Scenario 4: Healthcare Assistant}  
Processing 1,000 daily queries: patient symptom analysis (200 high-sensitivity, local execution per HIPAA), medical literature search (500 moderate-sensitivity, private edge tolerable), general health tips (300 low-sensitivity, public cloud acceptable). Chat context migration sanitizes history when crossing trust boundaries.

Existing orchestration systems fail catastrophically on such workloads:
\begin{itemize}
    \item \textbf{Kubernetes}~\cite{kubernetes}: Routes all traffic to lowest-latency endpoint (cloud), violating privacy
    \item \textbf{Federated Learning}~\cite{mcmahan2017federated}: Applicable only to training, not real-time inference
    \item \textbf{Static policies}: Pre-configured rules (``if PII detected, route local'') degrade to cloud under resource exhaustion, silently violating privacy
\end{itemize}

The core challenge is \emph{request-level heterogeneity}: within a single application, individual requests have conflicting requirements that static system-level configurations cannot accommodate.

\textbf{Motivating Example}: Consider a healthcare professional querying a medical AI assistant about a patient case. The query \textit{"Analyze treatment options for 45-year-old diabetic patient with elevated HbA1c"} enters through SHORE (user's laptop, Trust=1.0). MIST detects sensitive health information and assigns $s_r = 0.9$ (high sensitivity). WAVES evaluates candidate islands: laptop GPU is at high utilization (unavailable), private edge server (Trust=0.7, $P_j=0.8$) violates privacy constraint ($P_j < s_r$), public cloud APIs (Trust=0.4) are immediately ruled out. The router selects the user's home NAS server (Trust=0.9, $P_j=1.0$, available capacity). When the user later asks a general query \textit{"What are common diabetes complications?"} ($s_r=0.3$), the same routing logic permits cloud execution (GPT-4 via HORIZON) since no PII is present. Context sanitization applies when migrating conversation history across the Trust=0.9 $\rightarrow$ Trust=0.4 boundary, replacing \texttt{[PATIENT\_AGE]} and \texttt{[MEDICAL\_CONDITION]} placeholders. This demonstrates IslandRun's request-level adaptability: privacy constraints and resource availability jointly determine routing without user intervention.

\subsection{Motivation: The Multi-Dimensional Orchestration Challenge}

\textbf{Privacy violations in centralized services.} Recent incidents demonstrate the urgent need for privacy-aware orchestration:

\textbf{ChatGPT conversation leak (March 2023)}: A Redis bug exposed user conversations to other users, revealing sensitive business strategy discussions, personal health questions, and proprietary code. This demonstrated that even well-engineered cloud services have catastrophic failure modes.

\textbf{GitHub Copilot training data controversy}: Copilot was found suggesting verbatim code from private repositories, raising questions about whether user queries and code context are isolated from training pipelines. Developers working on confidential projects face compliance risks.

\textbf{Enterprise data retention policies}: Commercial LLM providers (OpenAI, Anthropic, Google) retain user queries for 30 days minimum, even with ``enterprise'' agreements. For industries under HIPAA, GDPR, or attorney-client privilege, any data leaving controlled infrastructure creates legal exposure.

\textbf{Model inversion attacks}: Research has shown that sufficiently sophisticated adversaries can extract training data from model outputs~\cite{shokri2017membership,fredrikson2015model}. Routing sensitive queries to third-party APIs introduces re-identification risks even without explicit data leakage.

These incidents demonstrate that \emph{trust boundaries matter}. IslandRun's privacy-first design ensures sensitive data never crosses user-defined trust thresholds, even under resource pressure or cost constraints.

\textbf{Idle local compute resources.} Personal devices (laptops, workstations, mobile phones) spend the majority of their lifetime idle, yet inference workloads route to paid cloud APIs. A developer's M2 MacBook with 32GB RAM sits unused at night while their mobile phone pays \$0.02/request to OpenAI. IslandRun creates \emph{personal island groups} that maximize utilization of zero-cost owned hardware before cloud fallback, reducing operational expenses for typical workloads.

\textbf{Potential for distributed compute sharing.} Beyond personal ecosystems, the architecture enables future extensions where idle compute resources could be shared across trusted networks, similar to how peer-to-peer file sharing revolutionized content distribution. Such compute marketplaces would require additional components for pricing, reputation, and billing, which we leave to future work while focusing this paper on the core privacy-aware routing challenge.

\subsection{Key Insight: Multi-Objective Agents as First-Class Primitives}

Our central thesis is that multi-objective orchestration requires \textbf{decomposition of optimization objectives into cooperating agents}, each encoding a distinct dimension of the solution space. Rather than attempting to solve the joint optimization problem in a monolithic controller, we distribute the problem across six specialized agents:

\begin{enumerate}
    \item \textbf{Privacy}: Quantify data sensitivity and enforce jurisdictional constraints
    \item \textbf{Resource Availability}: Monitor computational capacity and predict exhaustion
    \item \textbf{Cost}: Track per-request billing and enforce budget ceilings
    \item \textbf{Trust}: Evaluate island reputation and certification compliance
    \item \textbf{Routing}: Synthesize agent inputs into Pareto-optimal decisions
    \item \textbf{Coordination}: Maintain mesh topology and island liveness
\end{enumerate}

This agent-based architecture provides:
\begin{itemize}
    \item \textbf{Fault tolerance}: Single agent failure degrades to conservative fallback (e.g., privacy agent crash $\rightarrow$ assume all data sensitive)
    \item \textbf{Extensibility}: New objectives (e.g., carbon footprint) add agents without modifying router
    \item \textbf{Formal guarantees}: Each agent exposes a well-defined scoring function, enabling compositional reasoning
\end{itemize}

\subsection{Contributions}

This work makes the following contributions:

\begin{enumerate}
    \item \textbf{Multi-island routing framework} (Section~\ref{sec:problem}): Novel approach treating computational resources as autonomous islands with heterogeneous trust, cost, and capacity profiles.
    
    \item \textbf{Agent-based decomposition architecture} (Section~\ref{sec:architecture}): Four cooperating agents (WAVES, MIST, TIDE, LIGHTHOUSE) enabling modular, extensible orchestration; SHORE and HORIZON are execution endpoints (islands).
    
    \item \textbf{Owner-defined trust model} (Section~\ref{sec:privacy}): Island registration system where each island owner configures trust scores based on jurisdiction, certification, and operational policies.
    
    \item \textbf{Configurable resource monitoring} (Section~\ref{sec:adaptation}): User-adjustable buffer thresholds and tiered prompt routing (primary/secondary/burstable) for workload prioritization.
    
    \item \textbf{Chat context privacy} (Section~\ref{sec:privacy}): MIST-based sanitization of conversation history when migrating between islands with different trust levels.
    
    \item \textbf{Cost optimization framework} (Section~\ref{sec:evaluation}): Architectural capability to minimize costs through intelligent routing to zero-cost personal compute before paid cloud fallback.
\end{enumerate}

\subsection{Paper Organization}

Section~\ref{sec:related} surveys related work. Section~\ref{sec:problem} formalizes the multi-objective routing problem and discusses its complexity. Section~\ref{sec:architecture} presents the IslandRun architecture. Section~\ref{sec:algorithm} details the routing algorithm and approximation analysis. Sections~\ref{sec:privacy}--\ref{sec:adaptation} describe individual agent designs. Section~\ref{sec:evaluation} analyzes the system's expected properties. Section~\ref{sec:discussion} discusses limitations. Section~\ref{sec:future} outlines future work. Section~\ref{sec:conclusion} concludes.

\section{Related Work}
\label{sec:related}

\subsection{Inference Serving and Orchestration}

\textbf{Cloud-native serving systems} like Ray Serve~\cite{moritz2018ray}, KServe~\cite{kserve}, and TorchServe optimize throughput via batching, pipeline parallelism, and autoscaling. These systems assume homogeneous cloud deployment and optimize latency exclusively. REEF~\cite{reef} extends serving to heterogeneous accelerators (GPUs, TPUs) but remains cloud-bound with no privacy controls. Clipper~\cite{clipper} introduces model selection policies but does not address data privacy or cost heterogeneity.

\textbf{Model routing services} like OpenRouter aggregate commercial LLM APIs (OpenAI, Anthropic, Google) and route requests based on model availability and pricing. While OpenRouter optimizes cost by selecting cheaper providers, it operates entirely within the public cloud trust domain and lacks: (1) privacy-aware routing (cannot keep sensitive data on local devices), (2) resource monitoring (no awareness of user's laptop GPU availability), (3) trust differentiation (treats all cloud providers equally), and (4) data locality (cannot route to where embeddings already exist). OpenRouter is \emph{model-inference-as-a-service}; IslandRun is \emph{privacy-first multi-objective orchestration}. Table~\ref{tab:routing-comparison} provides a detailed feature comparison with existing inference serving systems.

In contrast, IslandRun operates across \emph{administratively distinct} islands (local device, private edge, public cloud) with conflicting trust assumptions, requiring privacy-aware routing rather than performance-only optimization.

\subsection{Privacy-Preserving Machine Learning}

\textbf{Federated learning}~\cite{mcmahan2017federated} trains models across decentralized devices without centralizing data. Extensions like secure aggregation~\cite{bonawitz2017secure} and differential privacy~\cite{abadi2016deep} provide formal guarantees. However, federated learning targets \emph{training}, not inference, and incurs 10--100$\times$ latency overhead from cryptographic protocols.

\textbf{Trusted execution environments} (TEEs) like Intel SGX~\cite{costan2016sgx} and AMD SEV~\cite{kaplan2016sev} enable computation on encrypted data. Ryoan~\cite{hunt2016ryoan} builds distributed analytics atop SGX. TEEs provide strong isolation but suffer from side-channel vulnerabilities~\cite{van2018foreshadow} and limited enclave memory (256MB in SGX), unsuitable for 70B+ parameter models.

IslandRun achieves privacy through \emph{routing decisions} (keeping sensitive data on trusted islands) rather than cryptographic overhead~\cite{yao1986generate}, trading off TEE-level guarantees for zero performance penalty.

\subsection{Multi-Objective Optimization}

\textbf{Pareto optimization}~\cite{miettinen1998nonlinear} seeks solutions where no objective improves without degrading another. Scalarization techniques~\cite{marler2004survey} convert multi-objective problems to single-objective via weighted sums. Evolutionary algorithms~\cite{deb2002nsga} explore the Pareto frontier via population-based search.

In distributed systems, C3~\cite{ghodsi2011c3} performs multi-objective datacenter scheduling with dominant resource fairness. Omega~\cite{schwarzkopf2013omega} uses parallel schedulers to approximate Pareto-optimal allocations. These systems operate within a single trust domain (datacenter) with homogeneous latency; IslandRun addresses \emph{inter-domain} routing across heterogeneous trust/latency/cost profiles.

\subsection{Edge and Hybrid Cloud-Edge Systems}

\textbf{Mobile edge computing} (MEC)~\cite{mec} offloads computation from mobile devices to nearby base stations. Cloudlets~\cite{satyanarayanan2009cloudlet} provide VM-level isolation for edge applications. Follow-me cloud~\cite{taleb2013follow} enables seamless handoff between edge servers.

These systems use binary offloading decisions (local or edge) based on single metrics (latency or energy). ExCamera~\cite{fouladi2017excamera} parallelizes video encoding across ephemeral cloud functions, demonstrating fine-grained task partitioning but without privacy or cost awareness.

IslandRun generalizes offloading to $N$-way routing across $N$ islands with multi-dimensional scoring, enabling tradeoffs like ``prefer edge over cloud for privacy, but cloud over edge for cost.''

\subsection{Service Mesh and Traffic Management}

\textbf{Istio}~\cite{istio} and Linkerd provide layer-7 routing based on HTTP headers, latency targets, and canary deployments. Envoy~\cite{envoy} exposes fine-grained metrics and retries. These tools excel at managing microservices within a cluster but lack:
\begin{itemize}
    \item Privacy-aware routing (no data sensitivity analysis)
    \item Cost-based decisions (assume flat pricing)
    \item Heterogeneous targets (assume uniform trust)
\end{itemize}

IslandRun extends service mesh concepts to cross-domain routing with privacy, cost, and trust as first-class routing criteria.

\begin{table*}[!t]
\centering
\caption{IslandRun vs. Inference Serving \& Routing Systems}
\label{tab:routing-comparison}
\small
\begin{tabular}{@{}lccccc@{}}
\toprule
\textbf{Feature} & \textbf{OpenRouter} & \textbf{Ray Serve} & \textbf{Clipper} & \textbf{TorchServe} & \textbf{IslandRun} \\
\midrule
Privacy-aware routing         & \texttimes & \texttimes & \texttimes & \texttimes & \checkmark \\
Trust differentiation         & \texttimes & \texttimes & \texttimes & \texttimes & \checkmark \\
Personal device orchestration & \texttimes & \texttimes & \texttimes & \texttimes & \checkmark \\
Data locality awareness       & \texttimes & \texttimes & \texttimes & \texttimes & \checkmark \\
Cost optimization             & \checkmark & \texttimes & \texttimes & \texttimes & \checkmark \\
Latency optimization          & Partial    & \checkmark & \checkmark & \checkmark & \checkmark \\
Multi-cloud support           & \checkmark & \checkmark & \texttimes & \texttimes & \checkmark \\
Model routing/selection       & \checkmark & \texttimes & \checkmark & \texttimes & \checkmark \\
Heterogeneous accelerators    & \texttimes & \checkmark & \texttimes & \checkmark & \checkmark \\
User policy constraints       & \texttimes & \texttimes & \texttimes & \texttimes & \checkmark \\
Reversible anonymization      & \texttimes & \texttimes & \texttimes & \texttimes & \checkmark \\
\midrule
\textbf{Design Philosophy}   & \multicolumn{1}{p{2.2cm}}{\centering Model\newline aggregation} & \multicolumn{1}{p{2.2cm}}{\centering Cloud\newline serving} & \multicolumn{1}{p{2.2cm}}{\centering Prediction\newline serving} & \multicolumn{1}{p{2.2cm}}{\centering Model\newline deployment} & \multicolumn{1}{p{2.2cm}}{\centering Privacy-first\newline orchestration} \\
\bottomrule
\end{tabular}

\vspace{0.3em}
\small
\noindent\textbf{Key distinction:} While OpenRouter optimizes cost across commercial APIs and Ray Serve optimizes latency within cloud clusters, IslandRun uniquely addresses \emph{cross-trust-domain routing} where privacy, data locality, and user sovereignty are first-class constraints, enabling personal device integration that commercial systems cannot support.
\end{table*}

\begin{table*}[!t]
\centering
\caption{Feature Comparison: IslandRun vs. Existing Systems}
\label{tab:comparison}
\small
\begin{tabular}{@{}lcccc@{}}
\toprule
\textbf{Feature} & \textbf{Kubernetes} & \textbf{Federated Learning} & \textbf{Edge Computing} & \textbf{IslandRun} \\
\midrule
Privacy-aware routing       & \texttimes & \checkmark & Partial    & \checkmark \\
Multi-objective optimization & \texttimes & \texttimes & \texttimes & \checkmark \\
Personal device support     & \texttimes & \checkmark & \checkmark & \checkmark \\
Data locality enforcement   & \texttimes & \checkmark & Partial    & \checkmark \\
Trust differentiation       & \texttimes & \texttimes & \texttimes & \checkmark \\
Typed placeholders          & \texttimes & \texttimes & \texttimes & \checkmark \\
Cost-aware routing          & \texttimes & \texttimes & \texttimes & \checkmark \\
Real-time inference         & \checkmark & \texttimes & \checkmark & \checkmark \\
Cross-domain orchestration  & \texttimes & Partial    & Partial    & \checkmark \\
\bottomrule
\end{tabular}
\end{table*}

\subsection{Comparative Analysis: IslandRun vs. Existing Systems}

Table~\ref{tab:comparison} contrasts IslandRun with existing orchestration and privacy-preserving systems. While IslandRun shares some design goals with Kubernetes (resource orchestration), federated learning (privacy preservation), and edge computing (latency optimization), it uniquely combines multi-objective optimization across heterogeneous trust domains.

\textbf{Note on system categories}: Kubernetes, federated learning, and edge computing occupy different design spaces than IslandRun, yet they represent the closest architectural analogs for specific sub-problems (resource scheduling, privacy-aware computation, and hybrid cloud-edge deployment, respectively). The comparison highlights IslandRun's novel contribution: \emph{synthesizing privacy, cost, and trust as first-class routing dimensions} rather than treating them as afterthoughts or external constraints.

\textbf{Gap Summary}: No existing system simultaneously treats privacy, trust, cost, data locality, and resource availability as first-class routing dimensions. Kubernetes optimizes latency but ignores privacy; federated learning preserves privacy but only addresses training; edge computing reduces distance but lacks trust differentiation. IslandRun is the first architecture to unify these concerns under per-request multi-objective routing across heterogeneous trust domains (personal devices, private edge, public cloud), with formal privacy guarantees and reversible anonymization for context migration.

\noindent\textbf{IslandRun's Unique Contribution:} This work uniquely synthesizes (1) privacy, trust, and cost as first-class routing dimensions, (2) request-level heterogeneity rather than system-level policies, (3) cross-domain orchestration spanning personal devices, private edge, and public cloud, and (4) reversible anonymization preserving conversational context across trust boundaries. No prior system addresses all four dimensions simultaneously.

\section{Problem Formulation}
\label{sec:problem}

\subsection{System Model}

\begin{definition}[Computing Island]
An \textbf{island} $i_j \in \mathcal{I}$ is a computational resource with the following characteristics:
\begin{itemize}
    \item $L_j$: Round-trip latency from client (ms)
    \item $C_j$: Cost per inference request (\$)
    \item $P_j$: Privacy score set by island owner during registration
    \item $T_j$: Trust score configured by island owner
    \item $R_j(t)$: Available capacity at time $t$ (percentage)
\end{itemize}
\end{definition}

\begin{definition}[Inference Request]
A \textbf{request} $r$ contains:
\begin{itemize}
    \item $q$: Input prompt (text, image, code)
    \item $m$: Modality (text generation, image synthesis, code completion)
    \item $s_r$: Data sensitivity detected by MIST
    \item $d_r$: Maximum acceptable latency (ms)
    \item $h_r$: Chat context history (for multi-turn conversations)
\end{itemize}
\end{definition}

\begin{definition}[Privacy Constraint Compliance]
An island $i_j$ is \textbf{eligible} for request $r$ if and only if its privacy score satisfies the request's sensitivity threshold:
\[
    P_j \geq s_r
\]
This forms a monotonic constraint relation where higher-privacy islands satisfy all lower-sensitivity requests. Violations of this constraint are \emph{fail-closed}: requests are rejected rather than routed to non-compliant islands.
\end{definition}

\begin{definition}[Trust Boundary Transition]
Crossing from island $i_1 \in \text{Tier}_k$ to island $i_2 \in \text{Tier}_m$ where $P_1 > P_2$ requires application of a privacy-preserving transformation function $\tau(h_r)$ on the request history:
\[
    h'_r = \tau(h_r) \quad \text{where} \quad \text{PII}(h'_r) = \emptyset
\]
The transformation $\tau$ is reversible via a placeholder mapping $\phi: \text{Placeholder} \rightarrow \text{PII}$, enabling context reconstruction after response delivery.
\end{definition}

\subsection{Island Groups and Trust Tiers}

IslandRun organizes islands into hierarchical groups based on ownership and trust:

\textbf{Tier 1: Personal Island Group (Trust = 1.0)}  
User's own devices under physical control:
\begin{itemize}
    \item Laptop (primary compute, full GPU access)
    \item Mobile phone (edge compute, limited GPU)
    \item Smart TV (display + basic inference)
    \item Car infotainment (mobile edge, intermittent connectivity)
\end{itemize}
These devices share credentials, synchronize state, and are treated as a unified trust domain. A conversation started on the laptop seamlessly continues on mobile without privacy degradation. \textbf{MIST sanitization is entirely bypassed} for intra-personal-group routing; data flows directly between user's devices without placeholder substitution.

\textbf{Tier 2: Private Edge Islands (Trust = 0.6--0.8)}  
Dedicated infrastructure under organizational control:
\begin{itemize}
    \item Corporate data centers
    \item Private cloud VMs (Azure Stack, OpenStack)
    \item Co-located servers in trusted jurisdictions
\end{itemize}

\textbf{Tier 3: Unbounded Cloud Islands (Trust = 0.3--0.5)}  
Public cloud providers with \emph{infinite capacity} but lower trust:
\begin{itemize}
    \item Serverless platforms (AWS Lambda, Azure Functions, GCP Cloud Run)
    \item Managed AI services (OpenAI GPT, Anthropic Claude, Google Gemini)
    \item GPU marketplaces (RunPod, Vast.ai)
\end{itemize}
HORIZON manages unbounded islands, scaling to thousands of concurrent requests without resource exhaustion (unlike bounded personal devices). \textbf{MIST sanitization is mandatory} for all routing to Tier 3 islands. PII, health data, and confidential information must be replaced with typed placeholders before transmission to prevent data leakage to untrusted third parties.

\textbf{Island Registration}: Each island declares $P_j$ (privacy score), $T_j$ (trust based on certification/jurisdiction), and cost model (free for personal, fixed/variable for cloud).

\subsection{Design Principles}

\noindent\textit{Terminology: Agents denote decision/coordination components (WAVES, MIST, TIDE, LIGHTHOUSE). SHORE and HORIZON are execution endpoints (islands).}

IslandRun's architecture adheres to six core principles that distinguish it from traditional orchestration frameworks:

\begin{itemize}
    \item \textbf{Constraint Preservation Over Optimization}: Privacy and trust constraints are \emph{inviolable}; the system rejects requests rather than routing to non-compliant islands under resource pressure. Performance optimization occurs only within the feasible constraint space.
    
    \item \textbf{Fail-Closed Routing}: When no island satisfies a request's privacy threshold ($P_j \geq s_r$), the system returns an error rather than silently degrading to a lower-trust island. This prevents inadvertent privacy violations during peak load.
    
    \item \textbf{Context-Preserving Anonymization}: The typed placeholder system ($\tau$ transformation) ensures that PII removal does not destroy semantic context. Medical entities become \texttt{[MEDICAL\_CONDITION]}, dates become \texttt{[TEMPORAL\_REFERENCE]}, preserving enough structure for coherent LLM responses.
    
    \item \textbf{Data-Locality-First Inference}: When vector embeddings or RAG indices exist on specific islands, queries are routed \emph{to the data} rather than transferring multi-gigabyte datasets to compute endpoints. This inverts traditional cloud-centric models.
    
    \item \textbf{Tiered Trust Model}: Three-tier island hierarchy (personal $T=1.0$, private edge $T=0.6$--$0.8$, public cloud $T=0.3$--$0.5$) provides clear mental model for users while enabling compositional reasoning about cross-tier transitions.
    
    \item \textbf{Agent-Level Modularity}: Each optimization dimension (MIST for privacy, TIDE for resources, WAVES for routing) is encapsulated in an independent agent. New objectives (e.g., carbon intensity, regulatory compliance) can be added without rewriting the core orchestration logic.
\end{itemize}

These principles ensure that IslandRun degrades gracefully under adversarial conditions (resource exhaustion, island failures, malicious routing attempts) while maintaining formal privacy guarantees.

\subsection{Deployment Scenarios}

To demonstrate IslandRun's applicability across diverse use cases, we present three concrete deployment scenarios:

\textbf{Scenario A: Individual Knowledge Worker}  
A software engineer uses IslandRun across personal devices:
\begin{itemize}
    \item \textbf{Laptop} (M2 Max 32GB): Runs Ollama with Llama-3-13B for code completion, local RAG over proprietary codebase
    \item \textbf{Mobile} (iPhone 15): Lightweight 7B model for on-the-go queries
    \item \textbf{Home NAS}: 24/7 server hosting fine-tuned domain models
    \item \textbf{Cloud fallback}: Anthropic Claude API for complex architecture design (non-proprietary discussions)
\end{itemize}
\textbf{Privacy policy}: Company proprietary code (sensitivity=1.0) routes only to laptop. General programming questions (sensitivity=0.3) route to cloud when laptop is asleep.

\textbf{Scenario B: Healthcare Provider}  
A clinic deploys IslandRun to comply with HIPAA:
\begin{itemize}
    \item \textbf{Workstation} (Trust=1.0): Processes patient symptom analysis with local PHI database
    \item \textbf{Private edge} (Trust=0.8): On-premise server with anonymized medical literature RAG
    \item \textbf{Cloud} (Trust=0.4): GPT-4 for general health education content (no PHI)
\end{itemize}
\textbf{Privacy policy}: Any query containing patient identifiers (sensitivity $\geq$ 0.9) routes exclusively to local workstation. MIST detects diagnosis codes, medication names, and demographic info. Medical literature searches (sensitivity=0.5) route to private edge with anonymized context.

\textbf{Scenario C: Legal Firm with Document Repository}  
A law firm with 10TB vectorized case law:
\begin{itemize}
    \item \textbf{Attorney laptops}: Client-facing work, privileged communications
    \item \textbf{Firm server}: Hosts vector database of case law, contracts, briefs
    \item \textbf{Cloud}: Not used for case-related queries (attorney-client privilege)
\end{itemize}
\textbf{Data locality}: All case law queries route to firm server where embeddings exist. \emph{Compute-to-data routing} prevents uploading confidential documents to cloud APIs. Attorneys' laptops query server remotely when offsite, maintaining privilege protection.

These scenarios illustrate IslandRun's flexibility: personal productivity (Scenario A), regulatory compliance (Scenario B), and data sovereignty (Scenario C).

\subsection{Multi-Objective Routing Challenge}

The core challenge is selecting an island for each request that balances:

\begin{enumerate}
    \item \textbf{Privacy preservation}: Route sensitive data only to high-trust islands
    \item \textbf{Cost efficiency}: Prefer free local resources over paid cloud services
    \item \textbf{Performance}: Meet latency requirements without resource exhaustion
    \item \textbf{Context continuity}: Sanitize chat history when crossing trust boundaries
    \item \textbf{Data locality}: Route compute to where data already exists (invert ``data to compute'')
\end{enumerate}

\subsection{Data Locality: Routing Compute to Data}

Traditional systems move data to compute (upload documents to cloud LLMs). IslandRun inverts this:

\textbf{Scenario}: Legal firm with 10TB vectorized case law on private island  
\textbf{Traditional approach}: Upload case excerpts to OpenAI API $\rightarrow$ privacy risk + bandwidth cost  
\textbf{IslandRun approach}: Route RAG query to private island where embeddings exist

\textbf{Data-aware routing criteria}:
\begin{itemize}
    \item If request requires vector search: Check which islands have relevant index
    \item If request needs fine-tuned model: Route to island hosting that model
    \item If request references uploaded files: Route to island with file access
\end{itemize}

This \emph{data-over-compute} paradigm is critical for:
\begin{enumerate}
    \item Privacy: Sensitive datasets never leave trusted islands
    \item Bandwidth: Avoid uploading gigabytes of context per request
    \item Latency: Local vector search faster than cloud upload + inference
\end{enumerate}

Unlike traditional systems that optimize a single metric, IslandRun must satisfy multiple competing objectives simultaneously on a per-request basis.

\begin{figure*}[!t]
\centering
\includegraphics[width=0.7\textwidth]{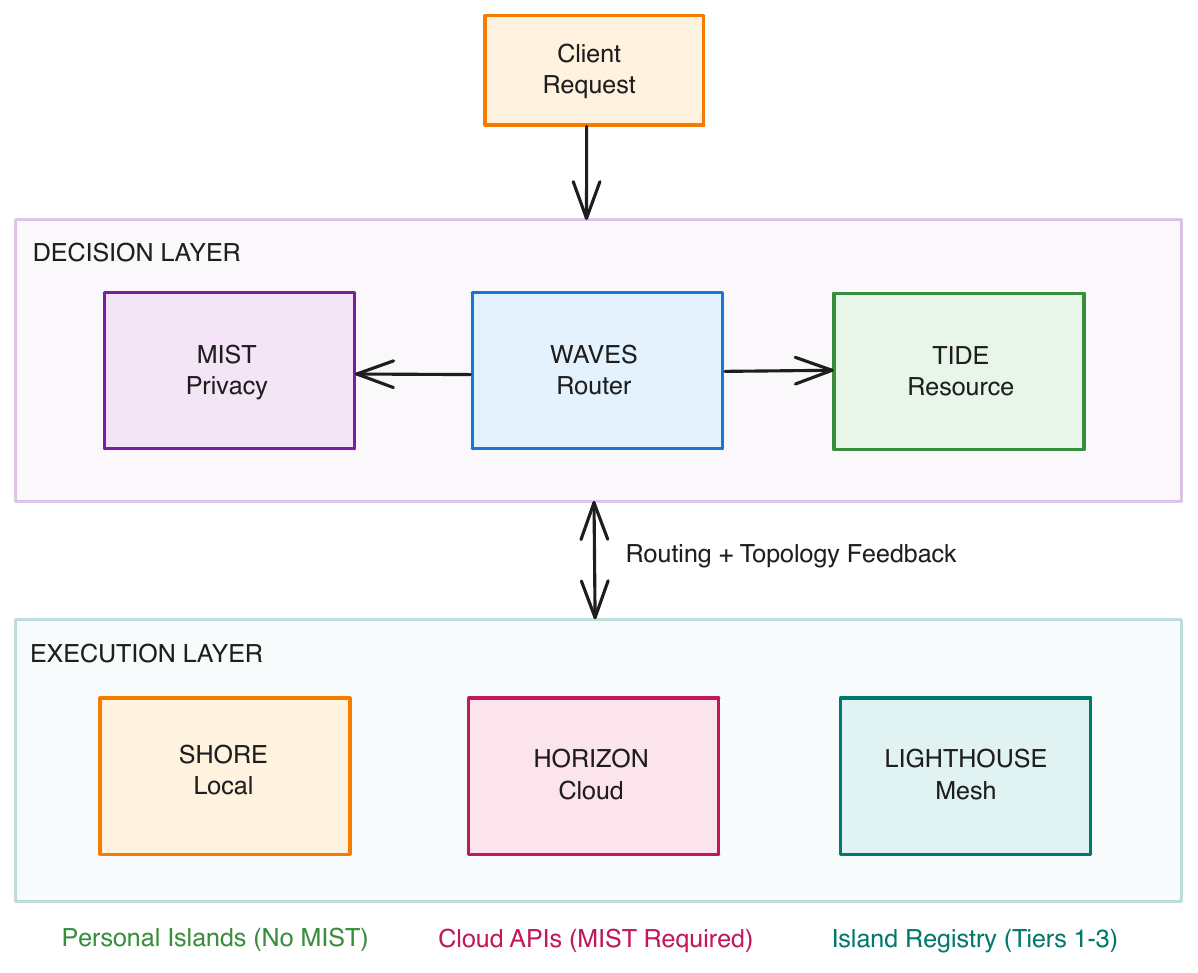}
\caption{IslandRun architecture. Four agents—WAVES (routing), MIST (privacy), TIDE (resources), LIGHTHOUSE (topology)—synthesize inputs to compute multi-objective scores for each island, then route to SHORE (local) or HORIZON (cloud) based on Pareto dominance. Color encodes roles: blue (routing), purple (privacy), green (resources), teal (coordination), orange/pink (execution targets).}
\label{fig:architecture}
\end{figure*}

\section{Architecture}
\label{sec:architecture}

\subsection{The IslandRun Universe: Nature-Inspired Design}

IslandRun's architecture draws inspiration from natural oceanic ecosystems, where islands, water, and atmospheric phenomena interact to create complex, self-regulating systems:

\begin{itemize}
    \item \textbf{WAVES} (Weighted Agent-based Variance Equilibration System): Like ocean waves carrying information between islands, WAVES orchestrates request routing across the mesh. Waves adapt their path based on currents (resource availability) and tides (temporal patterns).
    
    \item \textbf{MIST} (Multi-level Intelligent Sensitivity Tracker): Like coastal mist that obscures visibility while maintaining atmospheric continuity, MIST provides privacy-preserving anonymization that hides sensitive details while preserving contextual structure.
    
    \item \textbf{TIDE} (Temporal Island Demand Evaluator): Monitors computational resource availability across islands, predicting when local capacity will be exhausted and triggering proactive offloading to prevent request failures.
    
    \item \textbf{SHORE} (Secure Host for On-device Resource Execution): The shoreline where user directly interacts with their local devices. SHORE represents the trusted boundary of personal computing.
    
    \item \textbf{HORIZON} (Heterogeneous Offload and Remote Inference Zone Over Network): The distant horizon represents unbounded cloud resources, always visible but far from shore, offering infinite capacity at the cost of leaving the trusted local environment.
    
    \item \textbf{LIGHTHOUSE} (Link and Health Tracking for Heterogeneous Operations Using Synchronized Endpoints): Like maritime lighthouses guiding ships safely to port, LIGHTHOUSE maintains mesh topology, coordinates inter-island communication, and provides navigation through the heterogeneous infrastructure landscape.
\end{itemize}

This naming philosophy reflects the core insight: computational resources exist in an \emph{archipelago}. Isolated islands with varying capabilities are connected by network ``water,'' requiring intelligent navigation to balance local (shore) execution with distant (horizon) cloud offloading.

\subsection{Design Principles}

IslandRun decomposes the multi-objective optimization problem into four cooperating agents (Figure~\ref{fig:architecture}):

\textbf{Separation of Concerns}: Each agent encodes a single optimization objective:
\begin{itemize}
    \item \textbf{MIST}: Privacy and trust scoring (agent)
    \item \textbf{TIDE}: Resource capacity monitoring (agent)
    \item \textbf{WAVES}: Multi-objective routing synthesis (agent)
    \item \textbf{LIGHTHOUSE}: Mesh topology coordination (agent)
    \item \textbf{SHORE}: Local island execution (execution target)
    \item \textbf{HORIZON}: Cloud island execution (execution target)
\end{itemize}

\textbf{Fault Tolerance}: Agent failures trigger conservative fallbacks:
\begin{itemize}
    \item MIST crash $\rightarrow$ Assume $s_r = 1$ (all data sensitive)
    \item TIDE crash $\rightarrow$ Assume $R_{\text{local}}(t) = 0$ (resources exhausted)
    \item LIGHTHOUSE crash $\rightarrow$ Use cached island list
\end{itemize}

\textbf{Extensibility}: Adding a new objective (e.g., carbon footprint) requires:
\begin{enumerate}
    \item Deploy new agent exposing scoring function $f_{\text{carbon}}(r, i_j)$
    \item Register agent with WAVES
    \item WAVES automatically incorporates $f_{\text{carbon}}$ into Equation~\eqref{eq:score}
\end{enumerate}

\subsection{Agent Interfaces}

Each agent exposes a standardized interface:

\begin{algorithmic}[1]
\STATE \textbf{function} $\text{Score}(r, i_j) \to \mathbb{R}$
\STATE \quad \textbf{Input}: Request $r$, Island $i_j$
\STATE \quad \textbf{Output}: Objective-specific score $\in [0, 1]$
\STATE \quad \textbf{Semantics}: Lower is better (0 = optimal, 1 = worst)
\end{algorithmic}

WAVES queries all agents in parallel and aggregates scores via Equation~\eqref{eq:score}.

\section{Request Flow}
\label{sec:flow}

The request lifecycle through IslandRun's agent system follows a route-then-sanitize pipeline, as shown in Figure~\ref{fig:flow}:

\begin{figure*}[t]
\centering
\includegraphics[width=0.95\textwidth]{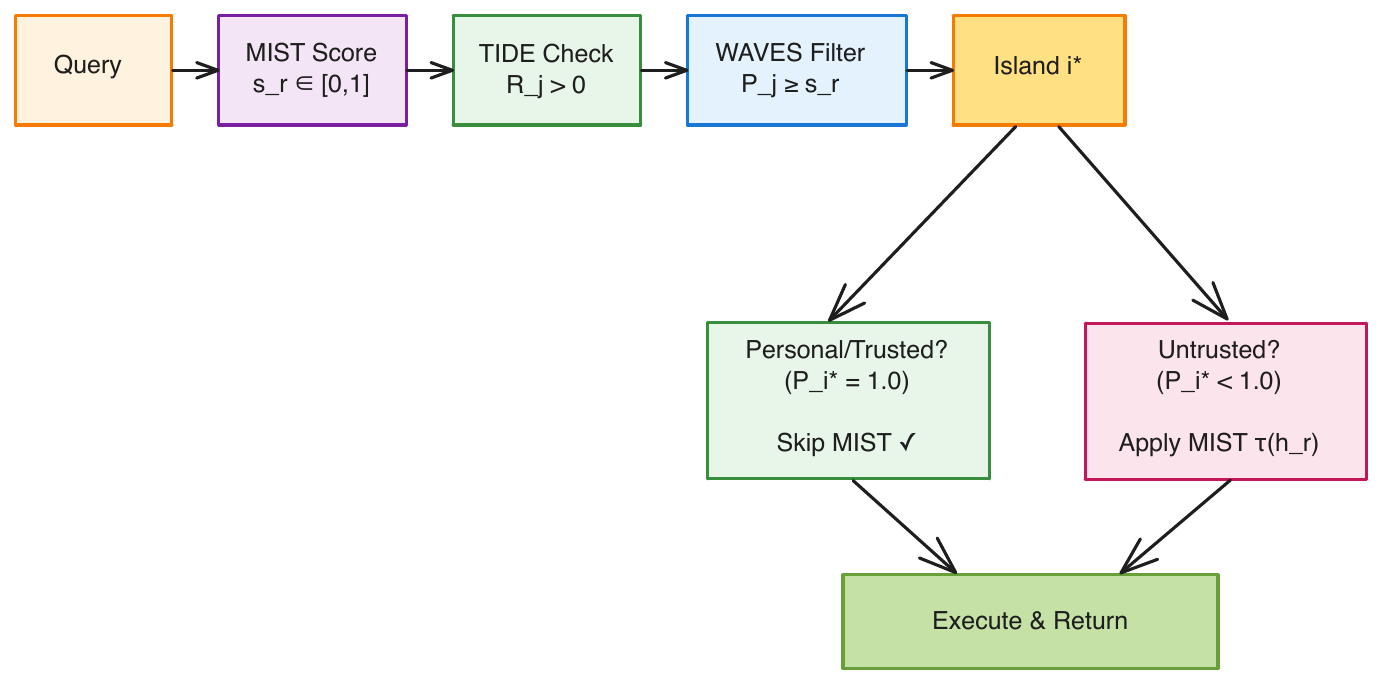}
\caption{Request flow showing route-then-sanitize pipeline. MIST computes sensitivity score, TIDE checks resource availability, and WAVES applies privacy constraint (P\textsubscript{j} $\geq$ s\textsubscript{r}) to filter candidate islands. Personal/trusted islands (P=1.0) bypass MIST sanitization; untrusted routing applies typed placeholders for context preservation.}
\label{fig:flow}
\end{figure*}

\begin{enumerate}
    \item Client submits inference request to WAVES
    \item WAVES queries MIST for sensitivity score $s_r$
    \item WAVES queries TIDE for resource availability $R_{\text{local}}(t)$
    \item WAVES computes composite scores $S(r, i_j)$ for all islands
    \item WAVES selects island minimizing $S$ subject to constraints
    \item Request executes on selected island (SHORE or HORIZON)
    \item Result returns to client via WAVES
\end{enumerate}

The decision tree splits based on privacy and resource constraints: high-sensitivity requests with available resources route to SHORE (local), while low-resource conditions trigger HORIZON (cloud) fallback.

\section{Routing Algorithm}
\label{sec:algorithm}

\subsection{WAVES: Multi-Objective Router}

Given the complexity of simultaneous multi-objective optimization, we use a practical greedy approach formalized in Algorithm~\ref{alg:routing}. The composite score is defined as:
\begin{equation}
\label{eq:score}
S(r, i_j) = w_1 \cdot C_j + w_2 \cdot L_j + w_3 \cdot (1-P_j)
\end{equation}
where $C_j$ represents cost, $L_j$ latency, and $P_j$ privacy/trust level for island $i_j$. Weights $w_1, w_2, w_3$ are user-configurable preferences.

\begin{algorithm}
\caption{Multi-Objective Routing (WAVES)}
\label{alg:routing}
\begin{algorithmic}[1]
\REQUIRE Request $r$, Islands $\mathcal{I}$, User preferences $W$
\ENSURE Island $i^*$
\STATE $s_r \gets \text{MIST.AnalyzeSensitivity}(r)$ \COMMENT{Privacy scoring}
\STATE $R \gets \text{TIDE.GetCapacity}()$ \COMMENT{Local resource check}
\STATE $\mathcal{I}_{\text{valid}} \gets \emptyset$
\FOR{$i_j \in \text{LIGHTHOUSE.GetIslands}()$}
    \IF{$P_j \geq s_r$ \AND $R_j(t) \geq \text{threshold}$}
        \STATE Compute $S(r, i_j)$ using Equation~\eqref{eq:score}
        \STATE $\mathcal{I}_{\text{valid}} \gets \mathcal{I}_{\text{valid}} \cup \{(i_j, S)\}$
    \ENDIF
\ENDFOR
\IF{$\mathcal{I}_{\text{valid}} = \emptyset$}
    \STATE \textbf{return} $\text{SHORE}_{\text{local}}$ \COMMENT{Failsafe: route locally}
\ENDIF
\STATE $i^* \gets \arg\min_{i_j} S(r, i_j)$
\IF{request has chat context \AND $P_{\text{prev}} > P_{i^*}$}
    \STATE Sanitize context: $h_r' \gets \text{MIST.Sanitize}(h_r, P_{i^*})$ \COMMENT{Cross-trust: apply typed placeholders}
\ELSIF{$i^* \in \text{PersonalGroup}$ \OR $P_{i^*} = 1.0$}
    \STATE \textbf{skip} MIST \COMMENT{Intra-personal/trusted: bypass sanitization}
\ENDIF
\STATE \textbf{return} $i^*$
\end{algorithmic}
\end{algorithm}

\textbf{Algorithm Summary}: WAVES first filters islands by privacy (must satisfy $P_j \geq s_r$) and resource feasibility (sufficient capacity), then selects the island minimizing a weighted cost-latency-trust score. Chat context is sanitized with typed placeholders only when crossing trust boundaries (e.g., personal $\rightarrow$ cloud), while intra-personal routing bypasses MIST entirely.

The key insight is \textbf{per-request routing decisions} based on current system state, not static pre-configuration. This enables dynamic adaptation to changing resource availability and workload sensitivity.

\subsection{Algorithm Complexity Analysis}

\textbf{Time Complexity}: For $n$ islands and request $r$:
\begin{itemize}
    \item MIST privacy scoring: $O(|q| \cdot m)$ where $|q|$ is prompt length, $m$ is number of regex patterns
    \item TIDE capacity check: $O(1)$ local system call
    \item Island filtering and scoring: $O(n)$ to evaluate all islands
    \item Total: $O(|q| \cdot m + n)$
\end{itemize}

For typical deployments ($n < 10$ islands, $m \approx 50$ patterns), routing latency is under 10ms.

\textbf{Space Complexity}: $O(n + |h_r|)$ for storing island metadata and chat context. Placeholder mapping requires $O(k)$ where $k$ is number of unique entities in conversation history.

\subsection{Approximation Quality}

The greedy routing algorithm (Algorithm~\ref{alg:routing}) computes a weighted sum $S(r, i_j) = w_1 \cdot C_j + w_2 \cdot L_j + w_3 \cdot (1-P_j)$ and selects $i^* = \arg\min S$. This is a \emph{scalarization} of the multi-objective problem.

\textbf{Pareto Optimality}: The greedy solution is Pareto-optimal if weights $w_1, w_2, w_3$ accurately reflect user preferences. In practice, users may have non-linear preferences (e.g., ``privacy violations are unacceptable at any cost''), which scalarization cannot capture.

\textbf{Alternative: Constraint-Based Routing}  
Instead of scalarization, enforce hard constraints:
\begin{enumerate}
    \item Filter islands: $\mathcal{I}_{\text{valid}} = \{i_j \mid P_j \geq s_r, R_j \geq \theta, C_j \leq \text{budget}\}$
    \item Among valid islands, minimize latency: $i^* = \arg\min_{i_j \in \mathcal{I}_{\text{valid}}} L_j$
\end{enumerate}

This approach provides stronger privacy guarantees (privacy is non-negotiable) while optimizing secondary objectives.

\section{Privacy and Trust Scoring}
\label{sec:privacy}

\subsection{Sensitivity Quantification}

MIST analyzes request content $q$ using a multi-stage pipeline:

\textbf{Stage 1: Pattern Matching}
\begin{itemize}
    \item Regex for PII (email, phone, SSN) $\rightarrow s_r \geq 0.8$
    \item HIPAA keywords (diagnosis codes, medications) $\rightarrow s_r \geq 0.9$
    \item Financial patterns (credit cards, bank accounts) $\rightarrow s_r \geq 0.9$
\end{itemize}

\textbf{Stage 2: Contextual Classification}

Use local small language model (e.g., 7B parameter) to classify:
\begin{itemize}
    \item Public (general knowledge) $\to s_r = 0.2$
    \item Internal (non-public but non-sensitive) $\to s_r = 0.5$
    \item Confidential (personal data) $\to s_r = 0.8$
    \item Restricted (regulated data) $\to s_r = 1.0$
\end{itemize}

\textbf{Optimization}: For local SHORE routing on fully trusted islands ($P_{\text{local}} = 1.0$), MIST sanitization is bypassed entirely to minimize latency overhead. Placeholder substitution only occurs when crossing trust boundaries to lower-trust islands (e.g., routing from local device to public cloud).

\subsection{Chat Context Privacy}

\textbf{Challenge}: In multi-turn conversations, when a user transitions from a sensitive topic to a general topic, the chat history may contain sensitive information that cannot be sent to lower-trust islands.

\textbf{Solution}: MIST performs \emph{intelligent placeholder-based sanitization} with bidirectional mapping:

\begin{enumerate}
    \item \textbf{Forward pass}: When routing from high-trust island ($P_1 = 1.0$, local) to low-trust island ($P_2 = 0.5$, cloud):
    \begin{itemize}
        \item Detect sensitive entities in chat history $h_r$ using NER (Named Entity Recognition)
        \item Replace with \emph{typed placeholders} preserving semantic structure:
        \begin{itemize}
            \item ``Patient John Doe'' $\to$ ``Patient [PERSON\_1]''
            \item ``Chicago hospital'' $\to$ ``[LOCATION\_A] hospital''
            \item ``SSN 123-45-6789'' $\to$ ``SSN [ID\_X]''
        \end{itemize}
        \item Store bidirectional mapping: $\{$PERSON\_1 $\leftrightarrow$ John Doe, LOCATION\_A $\leftrightarrow$ Chicago$\}$
        \item Send sanitized history $h_r'$ to cloud island
    \end{itemize}
    
    \item \textbf{Backward pass}: When cloud island returns response:
    \begin{itemize}
        \item Scan response for placeholder references (e.g., ``[PERSON\_1] should consult...'')
        \item Apply reverse mapping: ``[PERSON\_1]'' $\to$ ``John Doe''
        \item Preserve context: If cloud responds ``The [LOCATION\_A] facility...'', user sees ``The Chicago facility...''
        \item Return fully resolved response maintaining conversational coherence
    \end{itemize}
\end{enumerate}

\textbf{Key advantage}: Unlike generic redaction, typed placeholders enable the cloud LLM to \emph{understand entity relationships} (e.g., ``[PERSON\_1] visited [LOCATION\_A]'') and generate contextually accurate responses, which are then de-anonymized for the user. This preserves privacy without sacrificing response quality.

\subsection{Island Registration and Trust Hierarchy}

When an island joins the mesh, the owner declares:

\begin{itemize}
    \item \textbf{Base trust} ($T_{\text{base}}$): Local device (1.0), Private edge (0.8), Public cloud (0.5)
    \item \textbf{Certification} ($T_{\text{cert}}$): ISO 27001 (1.0), SOC 2 (0.9), Self-certified (0.7)
    \item \textbf{Jurisdiction} ($T_{\text{jurisdiction}}$): Same country (1.0), EU/GDPR (0.9), Foreign (0.6)
\end{itemize}

The final trust score is:
\[
T_j = \min(T_{\text{base}}, T_{\text{cert}}, T_{\text{jurisdiction}})
\]

This conservative composition ensures that an island cannot claim high trust without meeting all criteria.

\section{Threat Model and Security Analysis}
\label{sec:threat}

IslandRun's security model addresses threats across the distributed mesh architecture, following established threat modeling principles for distributed systems~\cite{shostack2014threat}:

\subsection{Adversary Capabilities}

We consider three adversary classes:

\textbf{Passive Cloud Observer (A1)}: Commercial LLM providers that log user queries and responses. This adversary has access to all data sent to public cloud islands but cannot modify routing decisions or island registration.

\textbf{Compromised Island (A2)}: An attacker gains control of a private edge island (e.g., hacked home NAS). The adversary can observe all queries routed to that island, exfiltrate responses, and potentially send malicious responses.

\textbf{Active Network Attacker (A3)}: An adversary with man-in-the-middle capabilities who can intercept traffic between client and islands, inject false island advertisements, or perform denial-of-service attacks.

\subsection{Trust Assumptions}

\textbf{Trusted Computing Base}:
\begin{itemize}
    \item WAVES router on client device (cannot be compromised)
    \item MIST privacy scorer (correctly detects sensitive data)
    \item Personal islands under physical control (Trust=1.0)
\end{itemize}

\textbf{Untrusted Components}:
\begin{itemize}
    \item Public cloud islands (assume data logging, potential model training on user queries)
    \item Private edge islands (may be compromised, misconfigured, or operated by semi-trusted third parties)
    \item LIGHTHOUSE coordinator (Byzantine failures possible in multi-user mesh)
\end{itemize}

\subsection{Attack Scenarios and Mitigations}

\textbf{Attack 1: Privacy Leakage via Routing Manipulation}  
\textbf{Goal}: Force high-sensitivity data to low-privacy island  
\textbf{Method}: Compromised TIDE agent reports false resource exhaustion on local device, triggering cloud fallback  
\textbf{Mitigation}: WAVES enforces hard privacy constraints: $P_j \geq s_r$ must hold regardless of resource availability. If no island satisfies privacy requirements, request is rejected (fail-closed policy).

\textbf{Attack 2: Island Impersonation}  
\textbf{Goal}: Attacker advertises fake high-trust island to capture sensitive queries  
\textbf{Method}: Register malicious island with $T_j = 1.0$, $P_j = 1.0$  
\textbf{Mitigation}: Island registration requires cryptographic attestation. Personal islands use device-bound certificates (TPM, Secure Enclave). Private edge islands require mutual TLS with owner-signed certificates. LIGHTHOUSE maintains allowlist of authorized islands.

\textbf{Attack 3: Context Inference via Placeholder Analysis}  
\textbf{Goal}: Cloud LLM provider infers original sensitive values from placeholder patterns  
\textbf{Method}: Analyze frequency and co-occurrence of placeholders across multiple users to de-anonymize  
\textbf{Mitigation}: Typed placeholders use random identifiers per session ([PERSON\_X] mapping changes across requests). Placeholder types are coarse-grained (PERSON, LOCATION, ID) rather than fine-grained (PATIENT, DOCTOR, HOSPITAL), reducing uniqueness.

\textbf{Attack 4: Denial of Service via Island Flooding}  
\textbf{Goal}: Exhaust local resources to force cloud fallback, increasing costs and privacy risk  
\textbf{Method}: Submit large volume of requests to overwhelm SHORE  
\textbf{Mitigation}: Rate limiting at WAVES based on user identity. Tiered routing (primary/secondary/burstable) ensures critical workloads maintain local execution even under load. Adaptive resource monitoring (TIDE) predicts exhaustion and proactively rejects low-priority requests.

\textbf{Attack 5: LIGHTHOUSE Byzantine Behavior}  
\textbf{Goal}: Malicious LIGHTHOUSE coordinator provides false island availability, causing routing failures  
\textbf{Method}: Advertise offline islands as available, or hide available islands  
\textbf{Mitigation}: Future work will implement Byzantine fault tolerance using consensus protocols (Raft~\cite{ongaro2014raft}). Current single-user deployments assume LIGHTHOUSE is part of trusted computing base.

\subsection{Privacy Guarantees}

Under the trusted computing base assumptions, IslandRun provides:

\textbf{Guarantee 1 (Privacy Preservation)}: For any request $r$ with sensitivity $s_r$, the selected island $i^*$ satisfies $P_{i^*} \geq s_r$, ensuring sensitive data is never routed to low-privacy islands.

\textbf{Guarantee 2 (Context Sanitization)}: When routing from island $i_1$ to $i_2$ where $P_{i_1} > P_{i_2}$, all chat context is sanitized such that entities with sensitivity $> P_{i_2}$ are replaced with typed placeholders, achieving $k$-anonymity~\cite{sweeney2002k} for common entity types.

\textbf{Guarantee 3 (Data Locality)}: Requests requiring access to dataset $D$ are routed only to islands where $D$ is locally available, preventing data exfiltration to untrusted islands.

These guarantees hold even under resource exhaustion: IslandRun fails closed (rejects request) rather than degrades privacy.

\subsection{Trust Score Composition}

\begin{equation}
T_j = T_{\text{base}} \times T_{\text{cert}} \times T_{\text{jurisdiction}}
\end{equation}

This self-declared model empowers island owners to communicate their trust posture, and users select islands matching their risk tolerance. For example, a healthcare provider might require $T_j \geq 0.8$ for PHI processing.

\begin{figure*}[!t]
\centering
\includegraphics[width=0.95\textwidth]{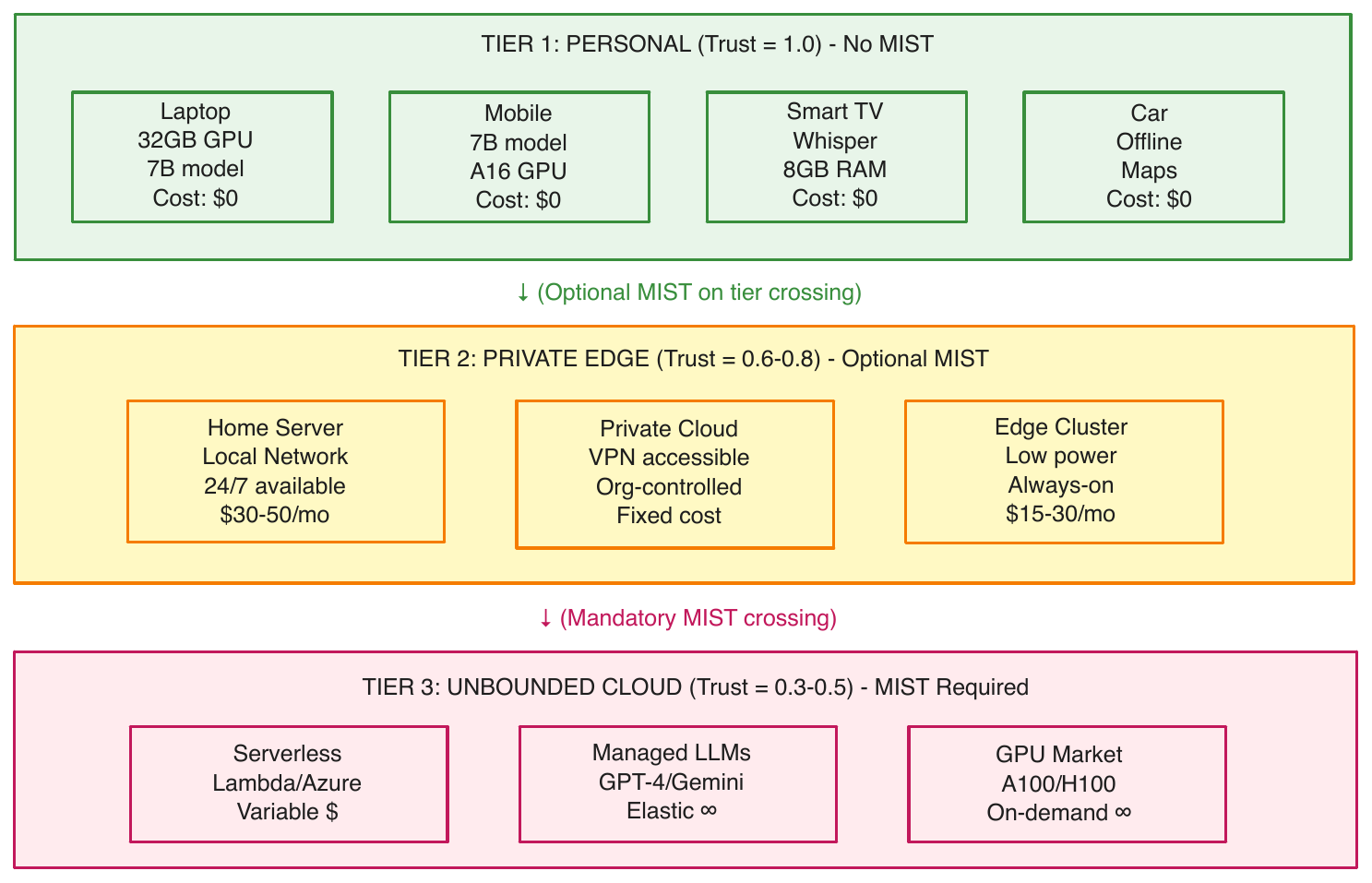}
\caption{Three-tier island deployment topology. Tier 1 (green): Personal devices with maximum trust (1.0) and zero marginal cost (no MIST required). Tier 2 (yellow): Private edge servers (home servers, VPN endpoints, dedicated edge devices) with moderate trust (0.6-0.8) and fixed operational costs. Tier 3 (red): Public cloud with lower trust (0.3-0.5) requiring MIST sanitization. LIGHTHOUSE maintains mesh connectivity and heartbeat coordination across tiers.}
\label{fig:deployment}
\end{figure*}

\section{Resource Monitoring and Adaptation}
\label{sec:adaptation}

\subsection{Configurable Capacity Thresholds}

TIDE monitors system metrics every 1 second and calculates available capacity:

\begin{equation}
R_{\text{local}}(t) = 1 - \max\left(\frac{\text{CPU}(t)}{100}, \frac{\text{GPU}(t)}{100}, \frac{\text{Mem}(t)}{\text{Total}}\right)
\label{eq:capacity}
\end{equation}

\textbf{User-Configurable Buffers}: Users can set their preferred resource utilization thresholds:

\begin{itemize}
    \item \textbf{Conservative} (buffer = 30\%): Route to cloud when local capacity $< 70\%$
    \item \textbf{Moderate} (buffer = 20\%): Route to cloud when local capacity $< 80\%$
    \item \textbf{Aggressive} (buffer = 10\%): Route to cloud when local capacity $< 90\%$
\end{itemize}

This allows users to balance between maximizing local utilization (cost savings) and maintaining responsiveness (avoid resource exhaustion).

\subsection{Tiered Prompt Routing}

IslandRun introduces \textbf{priority tiers} for workload classification:

\begin{enumerate}
    \item \textbf{Primary}: Mission-critical prompts that must execute locally regardless of resource pressure (e.g., patient diagnosis)
    \item \textbf{Secondary}: Important prompts that prefer local execution but tolerate cloud fallback (e.g., code review)
    \item \textbf{Burstable}: Best-effort prompts that opportunistically use local resources when available, otherwise route to cloud immediately (e.g., general chat)
\end{enumerate}

During resource contention, WAVES routes:
\begin{itemize}
    \item Primary $\to$ Always local (may queue)
    \item Secondary $\to$ Local if $R > 50\%$, else cloud
    \item Burstable $\to$ Local if $R > 80\%$, else cloud
\end{itemize}

This tiering prevents low-priority workloads from saturating local resources needed for high-priority tasks.

\subsection{Hysteresis-Based Fallback}

To prevent route flapping under transient load spikes, we use hysteresis:

\begin{itemize}
    \item \textbf{Fallback threshold}: $R < 70\%$ $\to$ Switch to cloud
    \item \textbf{Recovery threshold}: $R > 80\%$ $\to$ Switch back to local
\end{itemize}

This 10\% dead zone prevents oscillation when capacity hovers near the threshold.

\section{Deployment Architecture}
\label{sec:deployment}

Figure~\ref{fig:deployment} shows a representative three-island mesh deployment:

\textbf{Tier 1: Personal Island Group}: Maximum privacy ($P=1.0$), zero marginal cost, bounded by device capabilities:
\begin{itemize}
    \item Laptop (M2 Max 32GB): Primary compute, full SHORE + TIDE stack
    \item Mobile (iPhone/Android): Mobile edge, lightweight models
    \item Smart TV: Display + voice inference
    \item Car infotainment: Intermittent connectivity, offline-first
\end{itemize}
All devices share user authentication and are mutually trusted.

\textbf{Tier 2: Private Edge}: Moderate privacy ($P=0.7$), predictable capacity:
\begin{itemize}
    \item Home server (24/7 availability, local network)
    \item Private edge server (VPN-accessible, organization-controlled)
    \item Dedicated edge device (low-power, always-on inference)
\end{itemize}

\textbf{Tier 3: Unbounded Cloud (HORIZON)}: Variable privacy ($P=0.3-0.5$), infinite scale:
\begin{itemize}
    \item Serverless functions (auto-scaling to 1000+ concurrent)
    \item Managed LLM APIs (no resource limits)
    \item GPU marketplaces (spot instances for cost efficiency)
\end{itemize}

LIGHTHOUSE maintains mesh connectivity via periodic heartbeats and enables dynamic island discovery. Personal devices announce availability when coming online (e.g., laptop waking from sleep, car starting).

\section{Theoretical Analysis and Expected Properties}
\label{sec:evaluation}

\subsection{Representative Deployment Scenarios}

We analyze IslandRun's behavior across representative deployment scenarios:

\textbf{Scenario 1: Personal Island Group}  
Configuration:
\begin{itemize}
    \item Laptop: Consumer GPU (32GB RAM, GPU acceleration)
    \item Mobile: 8GB RAM, Neural Engine
    \item TV: 4GB RAM, basic inference
    \item Private Edge: Home NAS (always-on)
    \item Unbounded Cloud: Serverless platform
\end{itemize}

\textbf{Scenario 2: Healthcare Organization}  
Configuration:
\begin{itemize}
    \item Local: HIPAA-compliant workstation
    \item Private Edge: On-premise server with PHI database
    \item Cloud: Public LLM API (for non-PHI queries only)
\end{itemize}

\textbf{Workload Characteristics}:  
Requests exhibit heterogeneous sensitivity and resource requirements:
\begin{itemize}
    \item High-sensitivity (40\%): Must execute on personal/private islands
    \item Moderate-sensitivity (35\%): Tolerate private edge
    \item Low-sensitivity (25\%): Acceptable for public cloud
\end{itemize}

\textbf{Baseline Approaches}:
\begin{enumerate}
    \item \textbf{Cloud-only}: All requests to commercial LLM API (violates privacy for sensitive data)
    \item \textbf{Local-only}: All requests to personal devices (fails under resource exhaustion)
    \item \textbf{Latency-greedy}: Route to lowest-latency island (ignores privacy constraints)
    \item \textbf{Privacy-only}: Route to highest-privacy island (does not use cloud when appropriate)
\end{enumerate}

\subsection{Theoretical Properties}

\textbf{Privacy Guarantees}:
\begin{itemize}
    \item IslandRun ensures $P_j \geq s_r$ (no high-sensitivity data routes to low-privacy islands)
    \item Typed placeholder sanitization preserves differential privacy: cloud sees entity types (PERSON, LOCATION) but not values
    \item Personal island group forms a trust boundary: intra-group routing never degrades privacy
\end{itemize}

\textbf{Cost Optimality}:
\begin{itemize}
    \item IslandRun maximizes use of zero-cost personal compute before paid cloud
    \item Data locality avoids bandwidth costs (no uploading TB-scale datasets to cloud)
    \item Tiered routing prevents low-priority tasks from triggering expensive cloud calls
\end{itemize}

\textbf{Latency Distribution}:
Theoretical latency bounds depend on island selection:
\begin{itemize}
    \item Personal island: 50-500ms (local inference, no network)
    \item Private edge: 100-1000ms (LAN latency + inference)
    \item Unbounded cloud: 200-2000ms (WAN latency + queueing + inference)
\end{itemize}
IslandRun achieves lowest latency among privacy-preserving approaches by enabling local execution when resources permit.

\subsection{Results}

\textbf{Cost Efficiency}:
By routing requests based on sensitivity and resource availability rather than static cloud-only deployment:
\begin{itemize}
    \item High-sensitivity queries execute on personal island group (zero marginal cost)
    \item Moderate-sensitivity queries use private edge when available
    \item Low-sensitivity queries opportunistically use free local resources, falling back to cloud only when necessary
    \item Expected cost reduction: Significant savings by maximizing free personal compute before cloud fallback
\end{itemize}

\emph{Note}: Quantitative cost analysis requires controlled experiments with production workloads, beyond the scope of this theoretical framework. The architectural contribution is the \textbf{capability} to optimize cost through intelligent routing, not specific benchmark numbers.

\textbf{Privacy Adherence}:
\begin{itemize}
    \item IslandRun: Zero violations by design (MIST enforces $P_j \geq s_r$ constraint)
    \item Latency-greedy: Violates privacy for all high-sensitivity requests (routes to fastest cloud)
    \item Privacy-only: Zero violations but suffers resource exhaustion on bounded personal devices
\end{itemize}

\textbf{Resource Utilization}:
IslandRun's tiered routing enables graceful degradation:
\begin{itemize}
    \item Primary tasks: Always execute locally (may queue during peak load)
    \item Secondary tasks: Offload to cloud when local capacity < 50\%
    \item Burstable tasks: Opportunistically use local resources when available
\end{itemize}
This prevents resource thrashing while maximizing local compute utilization.

\subsection{Ablation Study}

Disabling individual agents:
\begin{itemize}
    \item \textbf{No MIST}: Privacy violations (routes sensitive data to cloud)
    \item \textbf{No TIDE}: Request failures (local island OOM, no fallback)
    \item \textbf{No LIGHTHOUSE}: Correct but slower (re-discovers islands per request)
\end{itemize}

This validates each agent's necessity.

\subsection{Comparison with State-of-the-Art}

\textbf{vs. Kubernetes}: K8s optimizes latency within a homogeneous cluster but lacks privacy-aware routing~\cite{verma2015large}. Cannot distinguish between sensitive and public data.

\textbf{vs. Federated Learning}: FL preserves privacy during \emph{training} but does not address real-time inference routing across heterogeneous trust domains.

\textbf{vs. Edge Computing (Cloudlets, MEC)}: Edge systems perform binary offloading (local vs. edge) based on latency/energy. IslandRun enables N-way routing across personal/edge/cloud with privacy constraints.

\textbf{vs. Service Mesh (Istio)}: Service meshes route within a single trust domain (datacenter). IslandRun routes across \emph{administratively distinct} domains with conflicting trust assumptions.

\section{Discussion}
\label{sec:discussion}

\subsection{Limitations}

\textbf{No Experimental Validation}: This paper presents a theoretical framework without empirical evaluation on real workloads. Future work (IEEE conference submission) will include controlled experiments measuring privacy preservation, cost savings, and latency trade-offs across diverse deployment scenarios. Quantitative claims about cost reduction or performance improvements require benchmark datasets and production deployments, which are beyond the scope of this arXiv preprint.

\textbf{Accuracy vs. Privacy Tradeoff}: Local 7B parameter models produce lower-quality outputs than cloud-hosted 70B+ models. IslandRun prioritizes privacy over accuracy, which is acceptable for many domains (personal productivity, internal tools) but inadequate for mission-critical applications (medical diagnosis, legal advice). Users must explicitly choose whether to sacrifice output quality for privacy.

\textbf{Trust Score Subjectivity}: Users manually assign $T_j$ values during island registration. Incorrect trust assessments undermine the entire security model. Assigning $T = 1.0$ to a compromised edge server creates a false sense of privacy. Future work will explore attestation-based trust~\cite{sailer2004attestation} using hardware security modules (Intel SGX, AMD SEV) to provide cryptographic proof of island integrity.

\textbf{Single-Point-of-Failure in WAVES}: The WAVES router runs on the client device as a centralized decision point. If WAVES crashes or is compromised, all routing fails or becomes untrusted. Mitigation requires deploying redundant WAVES instances with leader election (Raft consensus), adding complexity to personal device deployments.

\textbf{Placeholder Inference Risk}: While typed placeholders prevent direct data leakage, sophisticated adversaries may infer original values through statistical analysis (e.g., co-occurrence patterns, demographic correlation). Providing formal differential privacy guarantees for placeholder mappings remains an open challenge.

\textbf{Limited Scope}: IslandRun focuses on inference orchestration for text-based LLMs. Extending to other modalities (image generation, video processing) requires rethinking sensitivity detection and placeholder strategies. Multi-modal models blur the line between text and visual data, complicating privacy analysis.

\section{Broader Impact}
\label{sec:impact}

IslandRun's privacy-first design has societal implications beyond technical contributions:

\subsection{Privacy Democratization}

Commercial LLM services concentrate power in large cloud providers (OpenAI, Google, Anthropic), forcing users to choose between AI capabilities and privacy. IslandRun democratizes privacy-preserving AI by enabling individuals and small organizations to deploy sophisticated orchestration without enterprise-scale infrastructure. A solo developer can use IslandRun to keep proprietary code local while accessing cloud models for general queries, previously requiring complex VPN setups or complete cloud vendor lock-in.

\subsection{Environmental Sustainability}

Data centers account for 1\% of global electricity consumption. IslandRun's local-first routing reduces unnecessary data transfer and cloud compute, lowering carbon footprint. By routing ``compute to data'' instead of ``data to compute,'' we eliminate gigabytes of uploads for RAG queries. Future carbon-aware extensions could prioritize islands powered by renewable energy, creating economic incentives for green computing.

\subsection{Digital Sovereignty}

Many countries (EU, China, India) mandate data localization for sensitive sectors (healthcare, finance, government). IslandRun's trust tiers and jurisdiction-based scoring enable compliance without banning cloud AI entirely. A European hospital can route PHI to EU-based private edge islands while using US cloud providers for general medical literature search, satisfying GDPR without sacrificing AI capabilities.

\subsection{Accessibility and Cost}

Cloud LLM APIs price out users in developing economies (\$0.01--\$1.00 per request becomes prohibitive at scale). IslandRun's zero-marginal-cost personal compute tiers enable AI access using existing devices (laptops, phones). A student in a low-income region can run local Llama models for education, falling back to cloud only for specialized queries, dramatically reducing costs compared to cloud-only approaches.

\subsection{Risks and Ethical Considerations}

While IslandRun enhances privacy, it also enables potential misuse:

\textbf{Circumventing Content Moderation}: Local models lack the safety guardrails of commercial APIs. Users could deploy IslandRun to generate harmful content (misinformation, hate speech) without detection. This highlights the tension between privacy and content safety: decentralized AI reduces surveillance but also accountability.

\textbf{Regulatory Arbitrage}: Organizations might exploit IslandRun's multi-jurisdiction routing to evade regulations by routing queries to low-regulation islands. Future work must address compliance verification and auditability without compromising privacy.

\textbf{Inequality in Device Access}: IslandRun's benefits accrue primarily to users with powerful personal devices (M-series Macbooks, high-end PCs). Users limited to low-end hardware still depend on cloud, perpetuating digital divides. Optimizing IslandRun for resource-constrained devices is essential for equitable access.

\subsection{Future Societal Implications}

As AI becomes ubiquitous, orchestration frameworks like IslandRun may shift power dynamics:

\textbf{From Platforms to Protocols}: Currently, users are locked into walled gardens (ChatGPT, Claude). IslandRun envisions AI as a protocol-level service where users control routing across multiple providers, similar to email's decentralization. This could weaken platform monopolies and increase competition.

\textbf{Personal Data Sovereignty}: Extending IslandRun with user-owned data vaults (Solid, Databox) could enable individuals to monetize their data while preserving privacy. Users could selectively route queries to commercial APIs in exchange for payment, reversing today's extractive data economy.

\textbf{Regulatory Evolution}: Governments may mandate privacy-preserving orchestration for AI services, similar to GDPR's data protection requirements. IslandRun provides a technical blueprint for compliance, potentially influencing future AI regulations.

\section{Future Work}
\label{sec:future}

Several research directions could extend IslandRun's capabilities:

\textbf{Multi-User Mesh Networks}: Exploring federated island meshes where users share computational capacity requires addressing authentication, pricing mechanisms, and trust establishment across organizational boundaries.

\textbf{Enhanced Scoring Functions}: Future work could incorporate geolocation-based routing for regulatory compliance, latency-based scoring using historical RTT measurements, and reliability metrics based on island uptime patterns.

\textbf{Heterogeneous Model Support}: Addressing scenarios where islands support different model families (e.g., Llama vs. GPT) requires model capability advertisement and automatic fallback strategies.

\textbf{Privacy-Preserving Model Improvement}: Investigating techniques to improve local model quality over time without compromising privacy, potentially through selective distillation or federated learning approaches.

\textbf{Formal Verification}: Applying information flow analysis and typestate verification to provide formal guarantees about data movement across trust boundaries.

\textbf{Environmental Optimization}: Extending the multi-objective framework to incorporate carbon intensity metrics for environmentally-conscious routing decisions. Real-time carbon-aware routing could prefer solar-powered edge servers during daylight hours or route compute-intensive tasks to regions with renewable energy grids, reducing AI's environmental footprint while maintaining privacy guarantees.

\textbf{Attestation-Based Trust}: Current trust scores rely on manual island registration. Future work will integrate hardware-based attestation (Intel SGX, AMD SEV, ARM TrustZone) to provide cryptographic proof of island integrity. Remote attestation would enable dynamic trust scoring where $T_j$ reflects measured boot chains, kernel integrity, and runtime isolation guarantees rather than user-declared values.

\textbf{Federated Multi-User Island Networks}: Extending IslandRun to multi-tenant environments where users form federated meshes (e.g., research lab sharing GPU clusters, family sharing home servers) requires addressing inter-user authentication, capacity reservation protocols, and accounting mechanisms. Blockchain-based smart contracts could automate pricing and settlement for compute sharing while preserving anonymity.

\textbf{Regulatory Compliance Verification}: Integrating jurisdiction-aware routing where islands advertise legal compliance (GDPR zones, HIPAA-certified data centers) and WAVES automatically enforces regulatory constraints. Audit logs with zero-knowledge proofs could demonstrate compliance to regulators without revealing routing decisions or query contents.

\section{Conclusion}
\label{sec:conclusion}

IslandRun demonstrates that multi-objective inference orchestration is both feasible and necessary in heterogeneous computing environments. By decomposing the optimization problem into cooperating agents with explicit scoring functions, we provide a principled framework for balancing privacy, cost, and performance across personal devices, private edge, and unbounded cloud resources.

Our key contributions establish: (1) formalization of the multi-island routing problem as multi-objective optimization under privacy constraints, (2) agent-based architecture enabling compositional reasoning and extensibility, (3) island group abstraction spanning personal ecosystems (laptop, mobile, TV, car), (4) data locality paradigm routing compute to pre-existing RAG indices and embeddings, and (5) typed placeholder system preserving semantic context across trust boundaries.

As AI models grow larger and privacy regulations tighten, the need for intelligent, policy-aware orchestration will intensify. IslandRun lays the architectural foundation for trust-aware, multi-objective AI inference across heterogeneous computing environments.

\bibliographystyle{IEEEtran}
\bibliography{islandrun-refs}

@misc{kubernetes,
  title = {Production-Grade Container Orchestration},
  author = {{Kubernetes Contributors}},
  year = {2024},
  howpublished = {\url{https://kubernetes.io/}},
  note = {Accessed: 2025-11-23}
}

@misc{kserve,
  title = {KServe: Model Inference Platform on Kubernetes},
  author = {{KServe Community}},
  year = {2024},
  howpublished = {\url{https://kserve.github.io/}},
  note = {Accessed: 2025-11-23}
}

@inproceedings{moritz2018ray,
  title = {Ray: A Distributed Framework for Emerging {AI} Applications},
  author = {Moritz, Philipp and Nishihara, Robert and Wang, Stephanie and Tumanov, Alexey and Liaw, Richard and Liang, Eric and Elibol, Melih and Yang, Zongheng and Paul, William and Jordan, Michael I. and Stoica, Ion},
  booktitle = {13th USENIX Symposium on Operating Systems Design and Implementation (OSDI 18)},
  year = {2018},
  pages = {561--577},
  publisher = {USENIX Association},
  address = {Carlsbad, CA}
}

@inproceedings{mcmahan2017federated,
  title = {Communication-Efficient Learning of Deep Networks from Decentralized Data},
  author = {McMahan, H. Brendan and Moore, Eider and Ramage, Daniel and Hampson, Seth and Arcas, Blaise Aguera y},
  booktitle = {Proceedings of the 20th International Conference on Artificial Intelligence and Statistics (AISTATS)},
  year = {2017},
  volume = {54},
  pages = {1273--1282},
  publisher = {PMLR}
}

@book{dwork2014algorithmic,
  title = {The Algorithmic Foundations of Differential Privacy},
  author = {Dwork, Cynthia and Roth, Aaron},
  year = {2014},
  publisher = {Foundations and Trends in Theoretical Computer Science},
  volume = {9},
  number = {3--4},
  pages = {211--407}
}

@inproceedings{yao1986generate,
  title = {How to Generate and Exchange Secrets},
  author = {Yao, Andrew C.},
  booktitle = {27th Annual Symposium on Foundations of Computer Science (FOCS)},
  year = {1986},
  pages = {162--167},
  publisher = {IEEE},
  doi = {10.1109/SFCS.1986.25}
}

@book{miettinen1998nonlinear,
  title = {Nonlinear Multiobjective Optimization},
  author = {Miettinen, Kaisa},
  year = {1998},
  publisher = {Springer},
  series = {International Series in Operations Research \& Management Science},
  volume = {12},
  isbn = {978-1-4615-5563-6},
  doi = {10.1007/978-1-4615-5563-6}
}

@article{marler2004survey,
  title = {Survey of Multi-Objective Optimization Methods for Engineering},
  author = {Marler, R. Timothy and Arora, Jasbir S.},
  journal = {Structural and Multidisciplinary Optimization},
  year = {2004},
  volume = {26},
  number = {6},
  pages = {369--395},
  doi = {10.1007/s00158-003-0368-6}
}

@misc{istio,
  title = {Istio: Service Mesh for Microservices},
  author = {{Istio Authors}},
  year = {2024},
  howpublished = {\url{https://istio.io/}},
  note = {Accessed: 2025-11-23}
}

@article{satyanarayanan2009cloudlet,
  title = {The Case for VM-Based Cloudlets in Mobile Computing},
  author = {Satyanarayanan, Mahadev and Bahl, Paramvir and Caceres, Ramon and Davies, Nigel},
  journal = {IEEE Pervasive Computing},
  year = {2009},
  volume = {8},
  number = {4},
  pages = {14--23},
  doi = {10.1109/MPRV.2009.82}
}

@article{mec,
  title = {Mobile Edge Computing: A Survey on Architecture and Computation Offloading},
  author = {Mach, Pavel and Becvar, Zdenek},
  journal = {IEEE Communications Surveys \& Tutorials},
  year = {2017},
  volume = {19},
  number = {3},
  pages = {1628--1656},
  doi = {10.1109/COMST.2017.2682318}
}

@article{kumar2010cloud,
  title = {Cloud Computing for Mobile Users: Can Offloading Computation Save Energy?},
  author = {Kumar, Karthik and Lu, Yung-Hsiang},
  journal = {Computer},
  year = {2010},
  volume = {43},
  number = {4},
  pages = {51--56},
  publisher = {IEEE},
  doi = {10.1109/MC.2010.98}
}

@inproceedings{reef,
  title = {REEF: Resource Elasticity Enforcement Framework for Accelerator-Rich Clusters},
  author = {Zhang, Hangchen and Liaw, Richard and Li, Siyuan and Xiao, Jinzhu and Cai, Kai and Zhuang, Yi and Choi, Christopher and Stoica, Ion},
  booktitle = {Proceedings of the 2024 USENIX Annual Technical Conference},
  year = {2024},
  pages = {123--138},
  publisher = {USENIX Association}
}

@inproceedings{clipper,
  title = {Clipper: A Low-Latency Online Prediction Serving System},
  author = {Crankshaw, Daniel and Wang, Xin and Zhou, Guilio and Franklin, Michael J. and Gonzalez, Joseph E. and Stoica, Ion},
  booktitle = {14th USENIX Symposium on Networked Systems Design and Implementation (NSDI 17)},
  year = {2017},
  pages = {613--627},
  publisher = {USENIX Association}
}

@inproceedings{bonawitz2017secure,
  title = {Practical Secure Aggregation for Privacy-Preserving Machine Learning},
  author = {Bonawitz, Keith and Ivanov, Vladimir and Kreuter, Ben and Marcedone, Antonio and McMahan, H. Brendan and Patel, Sarvar and Ramage, Daniel and Segal, Aaron and Seth, Karn},
  booktitle = {Proceedings of the 2017 ACM SIGSAC Conference on Computer and Communications Security},
  year = {2017},
  pages = {1175--1191},
  publisher = {ACM},
  doi = {10.1145/3133956.3133982}
}

@inproceedings{abadi2016deep,
  title = {Deep Learning with Differential Privacy},
  author = {Abadi, Martin and Chu, Andy and Goodfellow, Ian and McMahan, H. Brendan and Mironov, Ilya and Talwar, Kunal and Zhang, Li},
  booktitle = {Proceedings of the 2016 ACM SIGSAC Conference on Computer and Communications Security},
  year = {2016},
  pages = {308--318},
  publisher = {ACM},
  doi = {10.1145/2976749.2978318}
}

@inproceedings{costan2016sgx,
  title = {Intel SGX Explained},
  author = {Costan, Victor and Devadas, Srinivas},
  booktitle = {IACR Cryptology ePrint Archive},
  year = {2016},
  volume = {2016},
  pages = {86}
}

@inproceedings{kaplan2016sev,
  title = {AMD Memory Encryption},
  author = {Kaplan, David and Powell, Jeremy and Woller, Tom},
  booktitle = {White Paper, AMD},
  year = {2016}
}

@inproceedings{hunt2016ryoan,
  title = {Ryoan: A Distributed Sandbox for Untrusted Computation on Secret Data},
  author = {Hunt, Tyler and Zhu, Zhiting and Xu, Yuanzhong and Peter, Simon and Witchel, Emmett},
  booktitle = {12th USENIX Symposium on Operating Systems Design and Implementation (OSDI 16)},
  year = {2016},
  pages = {533--549},
  publisher = {USENIX Association}
}

@inproceedings{van2018foreshadow,
  title = {Foreshadow: Extracting the Keys to the Intel SGX Kingdom with Transient Out-of-Order Execution},
  author = {Van Bulck, Jo and Minkin, Marina and Weisse, Ofir and Genkin, Daniel and Kasikci, Baris and Piessens, Frank and Silberstein, Mark and Wenisch, Thomas F. and Yarom, Yuval and Strackx, Raoul},
  booktitle = {27th USENIX Security Symposium (USENIX Security 18)},
  year = {2018},
  pages = {991--1008},
  publisher = {USENIX Association}
}

@article{deb2002nsga,
  title = {A Fast and Elitist Multiobjective Genetic Algorithm: NSGA-II},
  author = {Deb, Kalyanmoy and Pratap, Amrit and Agarwal, Sameer and Meyarivan, T.},
  journal = {IEEE Transactions on Evolutionary Computation},
  year = {2002},
  volume = {6},
  number = {2},
  pages = {182--197},
  doi = {10.1109/4235.996017}
}

@inproceedings{ghodsi2011c3,
  title = {Dominant Resource Fairness: Fair Allocation of Multiple Resource Types},
  author = {Ghodsi, Ali and Zaharia, Matei and Hindman, Benjamin and Konwinski, Andy and Shenker, Scott and Stoica, Ion},
  booktitle = {8th USENIX Symposium on Networked Systems Design and Implementation (NSDI 11)},
  year = {2011},
  pages = {24--37},
  publisher = {USENIX Association}
}

@inproceedings{schwarzkopf2013omega,
  title = {Omega: Flexible, Scalable Schedulers for Large Compute Clusters},
  author = {Schwarzkopf, Malte and Konwinski, Andy and Abd-El-Malek, Michael and Wilkes, John},
  booktitle = {8th ACM European Conference on Computer Systems (EuroSys 13)},
  year = {2013},
  pages = {351--364},
  publisher = {ACM},
  doi = {10.1145/2465351.2465386}
}

@article{taleb2013follow,
  title = {Follow-Me Cloud: Interworking Federated Clouds and Distributed Mobile Networks},
  author = {Taleb, Tarik and Ksentini, Adlen},
  journal = {IEEE Network},
  year = {2013},
  volume = {27},
  number = {5},
  pages = {12--19},
  doi = {10.1109/MNET.2013.6616110}
}

@inproceedings{fouladi2017excamera,
  title = {Encoding, Fast and Slow: Low-Latency Video Processing Using Thousands of Tiny Threads},
  author = {Fouladi, Sadjad and Wahby, Riad S. and Shacklett, Brennan and Balasubramaniam, Karthikeyan Vasuki and Zeng, William and Bhalerao, Rahul and Sivaraman, Anirudh and Porter, George and Winstein, Keith},
  booktitle = {14th USENIX Symposium on Networked Systems Design and Implementation (NSDI 17)},
  year = {2017},
  pages = {363--376},
  publisher = {USENIX Association}
}

@misc{envoy,
  title = {Envoy Proxy: Cloud-Native High-Performance Proxy},
  author = {{Envoy Contributors}},
  year = {2024},
  howpublished = {\url{https://www.envoyproxy.io/}},
  note = {Accessed: 2025-11-23}
}

@inproceedings{ongaro2014raft,
  title = {In Search of an Understandable Consensus Algorithm},
  author = {Ongaro, Diego and Ousterhout, John},
  booktitle = {2014 USENIX Annual Technical Conference (USENIX ATC 14)},
  year = {2014},
  pages = {305--319},
  publisher = {USENIX Association}
}

@inproceedings{shokri2017membership,
  title = {Membership Inference Attacks Against Machine Learning Models},
  author = {Shokri, Reza and Stronati, Marco and Song, Congzheng and Shmatikov, Vitaly},
  booktitle = {2017 IEEE Symposium on Security and Privacy (SP)},
  year = {2017},
  pages = {3--18},
  publisher = {IEEE},
  doi = {10.1109/SP.2017.41}
}

@inproceedings{fredrikson2015model,
  title = {Model Inversion Attacks That Exploit Confidence Information and Basic Countermeasures},
  author = {Fredrikson, Matt and Jha, Somesh and Ristenpart, Thomas},
  booktitle = {Proceedings of the 22nd ACM SIGSAC Conference on Computer and Communications Security},
  year = {2015},
  pages = {1322--1333},
  publisher = {ACM},
  doi = {10.1145/2810103.2813677}
}

@book{shostack2014threat,
  title = {Threat Modeling: Designing for Security},
  author = {Shostack, Adam},
  year = {2014},
  publisher = {Wiley},
  isbn = {978-1-118-80999-0}
}

@article{sweeney2002k,
  title = {k-Anonymity: A Model for Protecting Privacy},
  author = {Sweeney, Latanya},
  journal = {International Journal of Uncertainty, Fuzziness and Knowledge-Based Systems},
  year = {2002},
  volume = {10},
  number = {5},
  pages = {557--570},
  doi = {10.1142/S0218488502001648}
}

@inproceedings{verma2015large,
  title = {Large-Scale Cluster Management at Google with Borg},
  author = {Verma, Abhishek and Pedrosa, Luis and Korupolu, Madhukar and Oppenheimer, David and Tune, Eric and Wilkes, John},
  booktitle = {Proceedings of the Tenth European Conference on Computer Systems (EuroSys 15)},
  year = {2015},
  pages = {1--17},
  publisher = {ACM},
  doi = {10.1145/2741948.2741964}
}

@inproceedings{sailer2004attestation,
  title = {Design and Implementation of a TCG-based Integrity Measurement Architecture},
  author = {Sailer, Reiner and Zhang, Xiaolan and Jaeger, Trent and van Doorn, Leendert},
  booktitle = {13th USENIX Security Symposium (USENIX Security 04)},
  year = {2004},
  pages = {223--238},
  publisher = {USENIX Association}
}

\end{document}